\documentstyle[twoside,fleqn,epsfig,rotating,espcrc2]{article}
\newcommand{\AmS}{{\protect\the\textfont2
  A\kern-.1667em\lower.5ex\hbox{M}\kern-.125emS}}

\title{Experiments at Large Underground Detectors}

\author{Sergio Petrera \address{I.N.F.N. and Dipartimento di Fisica, Universit\`a dell'Aquila \\ 
        Via Vetoio, I-67010 Coppito-L'Aquila, Italy}}
       
\begin{document}

\begin{abstract}
Some of the topics  discussed  during the 1997 workshop on `Theoretical
and Phenomenological Aspects of Underground Physics'
are  briefly reviewed.
\end{abstract}

\maketitle

\section{INTRODUCTION}

 In this review talk
I will   very briefly  cover a  subset  
of the topics discussed   during this 
very interesting and fruitful  Workshop. 
The topics  
are: nucleon decay,   supermassive monopoles,
Neutrino Astrophysics and finally Cosmic Ray physics.

This is    a partial list.
I've left out   many important topics, where large underground 
(hereafter shortened to UG) detectors
play a major role.
In  particular I've left out  the  exciting fields of solar and atmospheric neutrinos,
whose `anomalies'  are stimulating interpretations in terms 
of neutrino oscillations. For these subjects, the reader can refer
to the review talks by F. von Feilitzsch and E. Kearns at this Workshop.
I'll  touch  only partly  the fascinating field of UHE Neutrino
Astrophysics. This topic is undoubtedly  a matter of deep UG experiments and is the major 
goal of large under-water or under-ice  Cherenkov detectors. In particular
the search for neutrinos from the   
most  luminous objects in  the Universe ({\it e.g.}, Active Galactic Nuclei)
is one of the most fascinating challenges of High Energy Astrophysics
in the next century. The talk by F. Halzen at this Workshop covers
extensively this subject.

There is a common feature among the detectors which I'm going to describe.
Each of them exploits some mixture (even if to a different extent)
of tracking and calorimetric capabilities at intermediate energies (MeV through
GeV range). 
For this reason they are
generally multi-purpose experiments, covering simultaneously different
fields.  This feature can be easily recognized in 
Table~\ref{tab:trackcalo} where
the main parameters of the currently operating large underground  
trackers and/or calorimeters as well as the covered physics topics are listed.

\begin{table*}[hbt]
\setlength{\tabcolsep}{1.5pc}
\tabcolsep=6pt
\newlength{\digitwidth} \settowidth{\digitwidth}{\rm 0}
\catcode`?=\active \def?{\kern\digitwidth}
\caption{Operating UG trackers and calorimeters}
\label{tab:trackcalo}
\begin{tabular*}{\textwidth}{@{}l@{\extracolsep{\fill}}ccccccl}
\hline
Detector & Rock & E$_{\mu}$ at & Det. & Sensitive & Sensitive 
 & Low en. & Physics \\
 & overburden &  threshold & technique & mass & hor. area &
 threshold & topics \\
 & (m.w.e) &  (TeV) & & (KTon) & (m$^2$) & (MeV) & \\
\hline
Baksan & 850 & 0.22 & Liq. scint. & 0.38 & 280 & $\sim$8 & CR-$\mu$, $\nu$-SN,\\
&&&&&&&$\nu$-atm, $\nu$-astr\\
MACRO & 3700 & 1.4 & Liq. scint. & 0.56 & 920 & 7 &${\cal M}$, CR-$\mu$,  \\
& & & + gas det. & & & &$\nu$-SN, $\nu$-atm, $\nu$-astr\\
LVD & 3700 &1.4 & Liq. scint. & 0.56 & 200 & 
3$\div$6 & CR-$\mu$, \\
& & & + gas det. &($\rightarrow$1.8)  & & &  $\nu$-SN \\
Soudan2 & 2100 & 0.7 & Gas det. & $\sim$1 & 120 & - &N-dec, CR-$\mu$, \\
&&&&&&&$\nu$-atm, $\nu$-astr  \\ 
Super-K & 2700 &1.0 & Water Ch. & 22.5$\div$32 & $\sim$1000 & 6.5 & N-dec, CR-$\mu$, $\nu$-sol,\\
&&&&&&&$\nu$-SN, $\nu$-atm, $\nu$-astr \\
\hline
\multicolumn{8}{@{}p{160mm}}{\footnotesize{Legenda: ``N-dec'', nucleon decay; 
``${\cal M}$'', monopole search; ``CR-$\mu$'',
Cosmic Ray muon flux and primary composition;
``$\nu$-sol'', solar neutrinos; ``$\nu$-SN'', neutrinos from Supernovae; 
``$\nu$-atm'', atmospheric neutrinos; ``$\nu$-astr'', neutrino 
astrophysics}}
\end{tabular*}
\end{table*}

\section{NUCLEON DECAY}

Historically nucleon decay originated the growth of UG physics in mid 
1970's. It 
was under the influence of Grand Unified Theories~\cite{GUT} that 
new physics topics claimed background free environments. The 
nucleon instability, probably the most exciting among the new phenomena
 predicted by these theories, stimulated several
 searches for nucleon decay in UG experiments~\cite{oldpdec}. Even if
 these experiments are usually considered to mark the birth of UG physics,
 nevertheless one must remember  that UG observations were already
 carried out for several years, mainly in CR and solar neutrino 
 physics. One of them in particular, the R. Davis $^{37}$Cl 
 experiment~\cite{Davis}, can be considered the father of all
 large UG detectors, starting data taking  since late 1960's.
 
 Various predictions from GUT's are given for several decay channels.
 The most favorable among them, p $\rightarrow$ e$^{+}$ $\pi^{0}$, can be
 taken as a benchmark for comparison among these theories. In minimal
 GUT, {\it i.e.} SU(5)~\cite{su5} , the lifetime to branching fraction ratio 
 ($\tau$/B) 
 for this channel is predicted to be
 4.5$\times$10$^{29\pm1.7}$ yr~\cite{su5epi0}. Larger symmetries, {\it e.g.}, 
 SO(10)~\cite{so10}, give
 higher values between 10$^{32}$ and 10$^{34}$ years. SUSY GUT's~\cite{susy} make
 new decay channels available and then new decay patterns to be 
 searched for. The channel  p $\rightarrow$ $\bar{\nu}$ $K^{+}$,
the dominant SUSY decay mode~\cite{pnuK}, is predicted
 in minimal SUSY (MSSM)~\cite{MSSM} to occur with $\tau$/B =
 10$^{34.5\pm1.2}$ yr.
 
 Experimental limits are currently available for various decay modes~\cite{PDG96}.
 The most stringent one refers to the above quoted p $\rightarrow$ e$^{+}$ $\pi^{0}$
 mode and is $\tau$/B $>$ 5.5$\times$10$^{32}$ yr~\cite{imb3} at 90\% C.L.. This value
 already rules out the minimal SU(5) GUT, but room is kept for larger symmetries.
 The SUSY preferred mode (p $\rightarrow$ $\bar{\nu}$ $K^{+}$)
is hard to be detected, mainly because of the unavoidable atmospheric
neutrino background. The best (background subtracted) limit has
been obtained by Kamiokande, $\tau$/B $>$ 1$\times$10$^{32}$ yr~\cite{kam}.

At present two experiments are searching for nucleon decays, Soudan~2 and
Super-Kamiokande (Super-K). The latter is already operational since 1996 and produced 
remarkable results on solar and atmospheric neutrinos. No results have
been yet reported on nucleon decay search, which is under analysis. The expectations
are anyhow very impressive, taking into account that the sensitive mass
has been increased roughly by a factor 10  and the track reconstruction
has been strongly enhanced with respect to the previous water
Cherenkov experiments (Kamiokande and IMB). If no signal is found, Super-K expects
to set new limits after 5 years of data taking at $\sim$10$^{34}$ yr and
$\sim$10$^{33}$ yr for the p $\rightarrow$ e$^{+}$ $\pi^{0}$ and the 
p $\rightarrow$ $\bar{\nu}$ $K^{+}$ modes respectively.

\begin{figure}[htb]
\begin{center}
\vspace{-1.5cm}
 \begin{sideways}
\mbox{
        \hspace{-0.26cm}\psfig{file=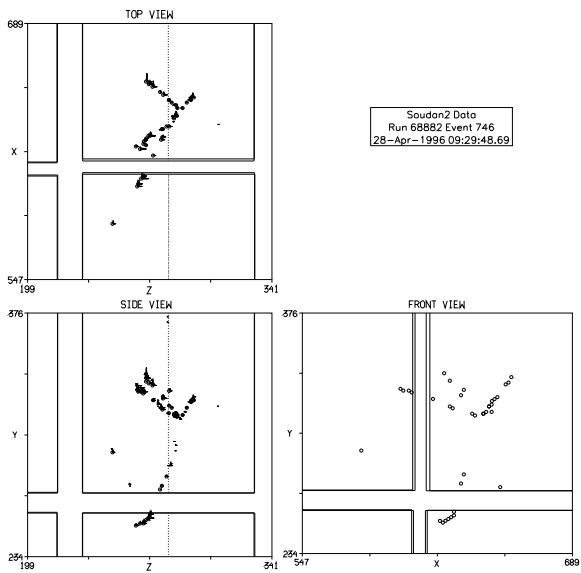,width=8.0cm}
     }
\
 \end{sideways}
\end{center}
\vspace{-1.4cm}
\caption{Candidate decay n $\rightarrow$ e$^{+}$ $\pi^{+}$ $\pi^{-}$ $\pi^{-}$ in Soudan 2.}
\label{fig:e3pi}
\end{figure}

New results on nucleon decay have been recently reported by the Soudan~2
Collaboration~\cite{soudan2icrc}. This preliminary analysis covers three decay classes:

\begin{itemize}
\item {\it $\bar{\nu}$ $K^{+}$ analysis}. In Soudan 2 this decay mode
is recognized identifying the kaon track (up to its stop), the
decay muon track and its subsequent decay. Kinematical cuts reduce the 
background to values 
not achievable with water Cherenkov techniques. After an exposure of 2.87 KTon~yr
no candidate has been found  with a background estimate of $\sim1$ event.
The overall detection efficiency times the branching fraction 
for this mode is $\sim$0.16. From this one gets $\tau$/B $>$ 3.5$\times$10$^{31}$ yr.
\item {\it 3-4 prong events}. A special effort has been dedicated to this class
of events for which Soudan 2 takes advantage from its good tracking capability
and vertex reconstruction ($\pm$1 cm). Many decay channels with this
topology are still uninvestigated (a complete list of the examined modes can be 
found in ref. \cite{soudan2icrc}). Intranuclear effects (rescattering or absorption of
 hadrons inside the nucleus) have been calculated with  Monte Carlo simulation.
 In an analysis  of 3.3 KTon~yr there are 12 of these events. All of them but one have
 far too much energy to be nucleon decay candidates and are consistent with
 neutrino multiparticle production. The only kinematically compatible
event is the n $\rightarrow$ e$^{+}$ $\pi^{+}$ $\pi^{-}$ $\pi^{-}$ candidate shown
in fig.~\ref{fig:e3pi}. Nevertheless this topology has inherently low probability to be
detectable ($\epsilon_{d} \sim 2\%$), on account of the high intranuclear absorption
probability.
\item {\it Exclusive decay modes}. In addition to the decay channels discussed above,
Soudan 2 has undertaken searches for the modes $\nu$$K^{0}_S$, $e^{+}$$K^{0}_S$,
$\mu^{+}$$K^{0}_S$, $\nu$$\eta$, $\nu$$\pi^{+}$, $\nu$$\pi^{0}$, $\nu$$e^{+}$$e^{-}$
$e^{+}$$\pi^{0}$, setting limits ranging from 0.2 through 0.9 10$^{32}$ yr.
\end{itemize}

\section{MAGNETIC MONOPOLES}

The search for  magnetic monopoles has been pursued for long time following the
Dirac paper~\cite{Dirac} in 1931. In
1974 t'Hooft~\cite{Hooft} and Polyakov~\cite{Polyakov}
showed that within the framework of Grand Unified Theories 
magnetic monopoles emerge naturally from the symmetry breaking
of the grand unified group into the strong and electroweak groups.
It is possible that this occurred in the early stages of the big
bang~\cite{Kolb}, producing a residue of primordial monopoles, for
which GUT's predict a mass of the order of $10^{16}$
to $10^{17}$~GeV/$c^2$. Relic monopoles are expected to have been cooled down
and now they are gravitationally trapped to the solar system or the
Galaxy. Under these circumstances their velocities relative to the Earth
range from $\beta\sim10^{-4}$ to $\beta\sim10^{-3}$, but acceleration 
mechanisms can be envisaged to allow for higher velocities. 
GUT's do not provide
any definite prediction about the  flux of magnetic monopoles. Astrophysical
arguments based upon the persistence of the interstellar
magnetic field~\cite{Parker} give an upper limit to this flux, $\Phi_{\cal M}$ 
$\leq$ $10^{-15}$~cm$^{-2}$~s$^{-1}$~sr$^{-1}$, usually referred to as the
``Parker Bound''. 
This low flux and the difficulty to detect slow moving magnetic particles
make this search possible only in large area underground detectors.
The only detector currently active in this field is MACRO. 

\begin{figure}[htb]
\begin{center}
\vspace{-1.5cm}
\mbox{
        \hspace{-0.26cm}\epsfig{file=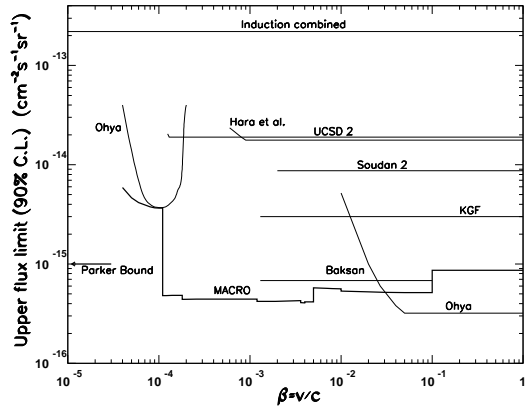,height=2.98in}
     }
\end{center}
\vspace{-1.4cm}
\caption{Magnetic monopole flux upper limits obtained by MACRO and by other 
experiments} 
\label{fig:worldlim}
\end{figure}

The MACRO 
detector is optimized to search for
GUT magnetic monopoles from $\beta\sim10^{-4}$ to $\beta=1$.  The
design goal is to reach a sensitivity an order
of magnitude below the Parker Bound for a five years' exposure.
Redundancy and complementarity among separate detector subsystems are a
central feature of the MACRO experiment. Three different techniques are 
used to reach this goal: i) He/n-pentane 
gas-filled streamer tubes; ii) liquid scintillator counters and
iii) nuclear track-etch detectors. Independent  
stand-alone ({\it i.e.} obtained with single subdetectors) analyses~\cite{cei} were 
already performed for the different $\beta$ intervals. The most recent 
analysis~\cite{cei}  gives a global MACRO limit , as the ``OR'' 
combination of the separate results. This flux limit is shown in 
fig.~\ref{fig:worldlim},
compared with upper limits by other experiments.
The MACRO limit has surpassed the 
Parker Bound by a factor 2 for $\beta > 10^{-4}$
and is the best existing for $10^{-4} < \beta < 5 \times 10^{-2}$.

The results obtained using the liquid scintillator and the track-etch
subdetectors can be, at least in part, extrapolated to the search for
nuclearites \cite{Witt84}. The flux upper 
limits are $\sim 10^{-15}$~cm$^{-2}$~s$^{-1}$~sr$^{-1}$ for nuclearites of mass 
$M_{N} > 0.1~{\rm g}$ and about $2$ times higher for $M_{N} < 0.1~{\rm g}$.

\section{NEUTRINO ASTROPHYSICS}

This is a very wide field which is approached with very different detection
methods and techniques. Large underground detectors are in particular active, and
 in some cases play a major role, in the following searches : i) $\nu$-emitting 
 point-like sources, ii) indirect search for weak interacting massive particles (WIMP's)
 and iii) neutrino bursts from Supernovae. The interest of neutrino
 astrophysics is addressed to many different objectives, which can be 
 relevant either to
 astrophysics, as in i) and iii), or particle physics, as in ii) and iii)
 ({\it e.g.}, properties of dark matter candidates or neutrinos).
 
 \subsection{Neutrino Astronomy}
 
 Because of their low cross-section, neutrinos are not absorbed at the creation sites
 and can reach us directly from the core of the astrophysical objects, bringing directly
  information about them. It is expected that astrophysical beam dumps which produce
$\gamma$-rays from $\pi^{0}$ decay, should also produce neutrinos from
$\pi^{\pm}$ decays. Possible high energy neutrino point-like sources are
X-ray binary systems, SN remnants, AGN's  and other sources with
significant $\gamma$ emission in the TeV range~\cite{gaisser95}.

UG experiments detect these neutrinos looking at upward-going or
nearly-horizontal muons. Muons from high energy neutrinos preserve their parent
direction (within few degrees). 
The restriction to zenith angles larger than 90$^\circ$
 allows one to detect neutrinos in a roughly background free
environment (the background being given by almost isotropical atmospheric
neutrinos).

MACRO reported a recent search~\cite{teresa} for neutrinos from several known sources.
No significant cluster of events has been found around any observable
direction 
and the largest signal found is of 2 events from Kepler 1604 (0.54 
are expected). The muon and neutrino flux limits for some sources
are given in Table~\ref{teresatab} and in some cases they improve previous limits already
given by Baksan~\cite{baksan95} and IMB~\cite{IMB95}. It has to be remarked
that these upper limits are one to two order of magnitudes higher than the expected
neutrino fluxes ({\it e.g.}, Gaisser~\cite{gaisser96} calculated for the SN remnant 
Vela Pulsar a
muon flux of the order of 0.03$\times$10$^{-14}$ cm$^{-2}$ s$^{-1}$ above $\sim$1 GeV).
A search for correlation with $\gamma$-ray bursts from BATSE catalogs 
is also reported in ref.~\cite{teresa}.
\begin{table*}[htb]
\setlength{\tabcolsep}{.5pc}
\caption{MACRO flux limits for some sources at $90\%$ c.l.
($\mu$-flux limits in $10^{-14}$ cm$^{-2}$ s$^{-1}$, $\nu$-flux limits in 
$10^{-5}$ cm$^{-2}$ s$^{-1}$). 
}
\label{teresatab}
\begin{tabular*}{\textwidth}{@{}l@{\extracolsep{\fill}}llllcl}
\hline
                 \multicolumn{1}{l}{Source} 
                 & \multicolumn{1}{l}{$\delta$} 
                 & \multicolumn{1}{l}{Events} 
                 & \multicolumn{1}{l}{Backg.}         
                 & \multicolumn{1}{l}{$\mu$-flux limits} 
                 & \multicolumn{1}{l}{Published $\mu$ limits} 
                 & \multicolumn{1}{l}{$\nu$-flux limits} \\
\hline
Cyg X-3&$40.6^{o}$&0&0.05&10.50&4.1 Baksan \protect\cite{baksan95}&4.25\\
MRK 421&$38.1^{o}$&0&0.07& 7.74&3.3 IMB \protect\cite{IMB95}&3.87\\
MRK 501&$38.45^{o}$&0&0.06&7.96& - &3.98\\
Crab Nebula&$22.0^{o}$&1&0.28&3.64&2.6 Baksan &1.82\\
Vela Pulsar&$-45.1^{o}$&0&0.86&0.61&0.78 IMB &0.30\\ 
Kepler 1604&$-21.3^{o}$&2&0.54&1.71&- &0.85\\ \hline
\end{tabular*}
\end{table*} 

Soudan 2 presented preliminary results~\cite{soudan2AGN} on a search for
high energy neutrinos from AGN's. The method follows  a previous search
from the Frejus experiment which led to a limit on diffuse neutrinos at
7$\times10^{-13}$ cm$^{-2}$ s$^{-1}$~\cite{frejus}.
 As UHE neutrinos from AGN's ($>$10 TeV) traverse
 the Earth, they can undergo charged-current interactions and produce muons. 
 These muons are likely to produce catastrophic energy loss by radiative processes,
 which dominate for energies in excess of 1 TeV. Neutrinos from AGN's are then
 searched looking at nearly horizontal muons exhibiting energy losses above 10 GeV.
 Efficiencies for measuring such energy deposition is calculated with Monte
 Carlo simulation at various muon energies. They observe no events and set a
 preliminary 90\% C.L. upper limit on diffuse  AGN $\nu$-induced muon flux  at the level of
 $10^{-13}$ cm$^{-2}$ s$^{-1}$ above 5 to 100 TeV of muon energy. This upper limit
rules out or at least severely constrains some of the AGN models (see {\it e.g.},
 ref.~\cite{gaisser96}).

\subsection{Indirect dark matter searches}

Deep UG muon detectors  can also look for  exotic  forms of 
dark matter, in particular WIMP's. 
The most plausible WIMP is considered the neutralino, $\chi$, the lightest
SUSY particle, which is  stable if R-parity is conserved.
The idea is the following:
WIMP's in the halo get gravitationally trapped and then accumulate  in  
the center  of the Sun, Earth
and other  astrophysical   bodies. Annihilation of these particles
with their anti-particles produce neutrinos of various
flavors, originating either from the decays of gauge or Higgs bosons or 
from the semileptonic decays of heavy quarks produced at annihilation.
 Some of these neutrinos  
are detectable as upward-going muons by a directional analysis. In practice
the only background to be faced is the one coming from atmospheric 
neutrinos\footnote{The signal  from the Sun  can 
 be observed   only at night,
when the Sun is  below the horizon.  Only  if the muons  are upward-going 
they can be  distinguished  from
the much larger flux of   atmospheric  muons produced in cosmic
ray  showers.}.

The angular spread between the parent neutrino and the detected muon 
is determined by the kinematics of the neutrino interaction, by multiple 
scattering of the muon from the interaction point to the detector and, for 
extended sources, by the dimensions of the annihilation region. In principle
this angle depends on the details of the final states of the annihilation,
but in practice the main parameter is the neutralino mass.

Data on upward muons from the Earth and the Sun have been measured by several 
experiments, notably Baksan~\cite{baksanwimps}, 
IMB~\cite{IMBwimps}, Frejus~\cite{frejuswimps}, Kamiokande~\cite{kamwimps}
and MACRO~\cite{MACROwimps}. The most recent analysis has been presented 
by MACRO~\cite{teresa} at this Workshop. The upper limit from the Sun
they reported is at the level of  $10^{-14}$ cm$^{-2}$ s$^{-1}$ for
$m_{\chi}\geq$ 30~GeV/c$^2$. This value has been compared with
muon fluxes obtained from neutrino fluxes calculated by Bottino et 
al.~\cite{bottino}, for various allowed MSSM parameters.

\subsection{Neutrinos from Supernovae}

Supernova explosions allows one to study the evolution end of very massive
stars from different points of view. It's still an ungranted dream
of physicists
to make combined observations of optical, neutrino and
gravitational wave radiations from a SN explosion. These  are
extraordinary events, at least in our Galaxy. The occurrence rate
depends on the detection method used: optically visible historical SN's 
have been seen at a rate of $\approx$ 0.5 century$^{-1}$; 
$\nu$-visible SN's are estimated at $\approx$ 2$\div$10 
century$^{-1}$~\cite{bemporad}.
Although these events are so rare and unpredictable,
the last ten years saw an unprecedented effort to build-up and
continuosly run  detectors capable to record
such events. In particular, a network of massive underground neutrino 
detectors is presently active, whose combined sensitivity to a galactic 
neutrino burst will  confirm and strongly improve the successful
observations~\cite{imbSN,kamSN} from the SN1987A explosion.

The $\nu$-radiation from SN's is detectable as a neutrino burst with
the following characteristics: initially there  is a short and  intense 
burst of  $\nu_e$,  due  to the `neutronization' of the star, then cooling
down is achieved through a  
longer `thermal emission' of almost `equipartitioned' neutrinos (among
the 6 flavors). The time scale of the neutronization burst is few  milliseconds,
the thermal burst lasts of the order of 10 sec.  
 As a whole of the order of 
10$^{57}$  $\nu$'s are emitted with an average energy of $\approx$ 10 MeV.
 
It's a difficult task for detectors to be sensitive to all of these
neutrinos. 
A complete analysis of the possible signals which are detectable in 
the existing underground experiments can be found in ref.~\cite{burrows}.
The detectors are of two types:
\begin{itemize}
\item {\it Liquid scintillator detectors}. These detectors 
(like Baksan, LSD, LVD and MACRO) use large masses 
($M\sim$ 1~KTon) of liquid scintillator, 
segmented in some hundred counters (observed by two or more 
PMT's for each) or enclosed in a container and observed by an array of 
PMT's at the boundary of the active volume. 
Good timing ($\sigma_t \sim 1~{\rm ns}$) and 
energy ($\sigma_E/E \sim 10\%$ at 10~MeV) resolutions are the most 
important features of these detectors. The segmented scintillator detectors 
have a good compatibility with tracking systems. This allows them
to reject the cosmic ray 
background.  
Liquid scintillator experiments are mainly sensitive to `thermal' 
$\bar{\nu_e}$'s by inverse beta reactions on protons.
The scintillation counters have generally a large light yield and 
  this makes the delayed secondary reaction $n + p \rightarrow d + \gamma$,
E$_\gamma$  = 2.2 MeV  detectable in these experiments.  
The $\gamma_{2.2}$ signal gives a powerful further signature of 
the $e^{+}$ event from the primary reaction. 
\item {\it Water/Heavy water Cherenkov detectors}. Experiments of this type
(like IMB and Kamiokande in the past, Super-K at present and SNO in
near future)  use large volumes of highly purified water (heavy water), 
equipped with an array of inward-looking phototubes to detect the Cherenkov 
light produced by relativistic charged particles.  The Cherenkov detectors 
have a continuous active medium and are self-shielded from the external 
radioactivity background. The energy threshold of these 
experiments is in the range 5 $\div$ 10~MeV. The two
new detectors (Super-K and SNO) can remarkably enrich the knowledge 
about the neutrino burst.
In particular Super-K for the first time can collect a sizeable amount of
$\nu_e$'s from the `neutronization' phase, whereas 
SNO will be sensitive to the neutral current interactions on 
Deuterium, thus allowing to detect neutrinos of different flavors.
\end{itemize}

No SN neutrino burst has been detected after 1987 by any of the operating
detectors. The analysis of the event clusters is compatible with the
background measured at the  different sites. More recently a special
effort is dedicated to provide experiments with online ``Early Warning'' 
systems. MACRO published a paper~\cite{SNwatch} showing its online
SN watch monitor.  The motivation for this system  derives from
the hope that a notification of a neutrino burst given within a 
short time ($\sim$one hour) 
increases the chance of observing the onset of the optical signal. 
Furthermore, coincident alarms emanating from more than one 
neutrino observatory merged into a  centralized computer repository offers 
enhanced sensitivity and directional information.  

\section{COSMIC RAYS}

\begin{figure}[htb]
\begin{center}
\mbox{
        \hspace{-0.26cm}\epsfig{file=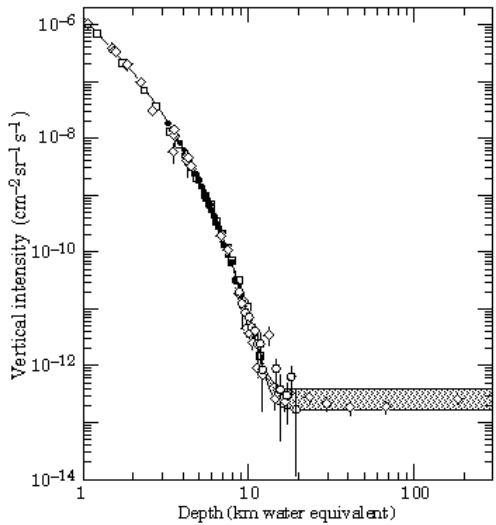,height=2.98in}
     }
\
\end{center}
\vspace{-1.4cm}
\caption{Vertical muon intensity {\it vs.} depth. The experimental data are
from: Crouch compilation (diamonds), Baksan (open squares), 
LVD (open circles), MACRO (full circles), Frejus (full squares). The shaded 
area at large depths represent neutrino-induced muons of energy above 2 GeV.
The upper line is for horizontal neutrino-induced muons, the lower one for
vertically upward muons.} 
\label{fig:vermuint}
\end{figure}

UG experiments study Cosmic Ray physics through the detection of 
the penetrating components of air showers (EAS).
 Only muons and neutrinos penetrate to significant
depths underground. Apart from neutrino detectors ({\it e.g.}, water Cherenkov or
fine-grained calorimeters) which are capable to identify GeV neutrino 
interactions in contained (or semi-contained) events, all  other detectors
measure only through-going muons, both atmospheric and neutrino-induced.
These measurements pertain to different
depth intervals: up to $\approx$10 Km.w.e. atmospheric muons are dominating,
at higher depths neutrinos constitute the only residual cosmic ray
component. This is illustrated in fig.~\ref{fig:vermuint} which shows the 
vertical muon intensity as a function of depth~\cite{Gaisserbook,PDG96}.

Large UG detectors collect copiously TeV muons, in a good fraction
grouped in muon bundles, and these are  used for two major
physics topics: i) study of primary CR spectrum and composition and 
ii) study of hadronic interaction mechanisms.  It is possible to decouple, 
at least partly,
 the two sources in large area
detectors: the main requirement is that their lateral sizes be large
with respect to the lateral spread of the muon bundles (typically of the order
of a few meters at the depths of UG detectors).
 
The measurement of the primary composition at high energies 
($\ge$ 100 TeV)
and of its possible variations around the steepening of the primary spectrum 
(the ``knee'', at about 2$\times$10$^{3}$ TeV), is
one of the main experimental problems in Cosmic Ray physics.
Due to low fluxes, measurements
must be indirect, {\it i.e.} through the study of the EAS components.
In particular, the analysis of muon events detected deep underground
is one of the most interesting
tools for the indirect study of primary composition,
since it can be shown that the
muon multiplicity, for a given energy threshold of muons, is sensitive to
both the energy and mass number of the primary particle~\cite{CRphysics}.
Measurements are in general sensitive not only to the
primary spectrum and composition, but also to the interaction
properties.

There have been two recent papers on primary composition studies with UG muons,
from Soudan~2~\cite{soudan2comp} and from MACRO~\cite{macrocomp}. 

The Soudan~2 Collaboration performs a standard multimuon analysis
by a comparison of the measured muon multiplicity distribution with
predictions from trial composition models. They use three compositions,
two of which are physically motivated by the assumption of a new CR source,
 as extension
of the basic supernova acceleration mechanism~\cite{Biermann}.
 The muon multiplicities
used for this analysis range from 6 to 12, roughly corresponding to
primaries between 8$\times$10$^2$ and  1.3$\times$10$^4$ TeV. They conclude
that, out
of the three compositions analysed, their data favor the lightest one
(lower average mass $<$A$>$). I would like to remark that the average mass
evolution of the three models is considerably different above the knee,
but the strongest difference, in the energy region
covered by this analysis, is between the heaviest composition and
the other two. Therefore I would prefer to interpret their results as
a definite inconsistency  with the
predictions of an asymptotically Fe-dominated composition.

The MACRO Collaboration has derived the chemical composition making use of
a best fit of the multimuon rates, based on five elemental
spectra described by two-power law functions and a rigidity dependent cutoff. 
The large MACRO
detector  acceptance and its good tracking capability allows them
  to perform a study of multiple muon events at 
high muon multiplicities (up to about 40 muons) and large
separations, essentially unaffected by finite detector size biases.
From their best fit analysis they estimate the primary composition parameters 
on a wide energy
interval (ranging from a few 10 TeV up to 10$^{5}$ TeV).
A remarkable feature of the reconstructed
all-particle spectrum, which derives from the fitting procedure, is
the sensitivity of MACRO data to the knee\footnote{For the first time 
it is shown that UG muons do ``see'' the knee. Previous UG
studies used composition models already incorporating the knee, but
didn't clearly showed requirement of it.}  
and a good consistency with EAS array measurements. However the
fitted spectrum is higher and flatter than the one obtained
from direct measurements ($\sim$15\% at 10 TeV up to $\sim$50\% at 100 TeV).
This  disagreement, as well as similar differences  in the TeV muon 
yields\footnote{Too few muons are predicted using low energy spectra
from direct measurements in the framework of currently used hadronic 
interaction models~\cite{Gaisser_vulcano,Fletcher,Ranft,Knapp}.}, may be due
 to possible inadequacies of the hadronic interaction model (see below).
In fig.~\ref{fig:avga_1} the average mass number is displayed as a function
of primary energy and compared with other measurements. 
$<$A$>$ shows little dependence on the primary energy below about 1000 TeV.
At higher energies the best fit average mass shows a mild increase
with energy, even though no definite conclusion can be reached
taking into account the increasingly large uncertainties deriving from
the fit.
%
\begin{figure}[htb]
\begin{center}
\mbox{
        \hspace{-0.26cm}\epsfig{file=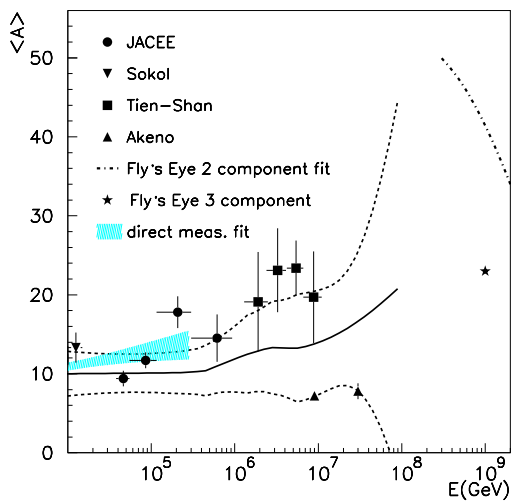,height=2.98in}
     }
\
\end{center}
\vspace{-1.4cm}
\caption{Average primary 
mass arising from MACRO fit (solid line: central value; dashed line:
value at one sigma error) compared  with other measurements.
$<$A$>$ is displayed
up to $\sim$ 10$^{9}$ GeV, exceeding the region covered by MACRO by
more than one decade, in order to include the composition results
from Fly's Eye in the EeV region.} 
\label{fig:avga_1}
\end{figure}
%

This study shows
how high statistics and good quality of data can provide enough
discrimination power to make a real composition measurement and not only a mere
comparison with trial models. Nevertheless, for this kind of analyses,
based on a single measured parameter (in this case muon multiplicity),
 the deconvolution of the primary spectrum from the experimental data is 
to some extent dependent on the particular nuclear interaction model used.
These models are built-up in such a way to reproduce available
experimental data, which are  anyhow limited in energy ($\sqrt s\leq$1 TeV, 
corresponding to a proton  Lab energy E$_p\leq$ 500 TeV)  and in the knowledge
of nuclear interaction mechanisms at high energies.
 Therefore one could believe
that possible inadequacies of interaction models are increasing with energy, but
are virtually absent in the energy region below the knee. 
A more careful study about the relevant
kinematical region (e.g., the Feynman-x interval) accessed by CR primaries 
producing UG muons shows that
possible uncertainties are also present at lower energies.
In particular one can see
that multimuon events originating from primaries below the knee are 
preferentially produced from parents in the very forward fragmentation region, 
yet very little data are available at x$_F$ exceeding 0.1.

A certain reduction of the dependence of the analysis on the
interaction model is achieved making multiple measurements of
at least two components of the EAS (one of them being usually
the electromagnetic component).
The discrimination power of the analysis is strongly enhanced
with respect to the previous approach and therefore the measured composition
is generally less dependent on EAS modeling through the air. 
This approach is followed in most of the CR studies from surface
detectors (EAS arrays, air Cherenkov, fluorescence detectors, generally
combined in the same site)~\cite{rappicrc95}. In few sites it is possible,
and it is indeed realized, to make observations combining UG muons and
surface EAS parameters: EAS-TOP/MACRO, EAS-TOP/LVD,
AMANDA/SPASE/VULCAN, Soudan-2/Air-Cherenkov~~ and ~~Baksan/Air-
Cherenkov.
Among these the experimental programs at Gran Sasso are probably the most
advanced. In particular, EAS-TOP/MACRO presented two recent analyses performed 
combining  the shower size N$_e$ at surface (from EAS-TOP) with the UG muon
multiplicity N$_\mu$ (from MACRO). Two classes of events are selected: high 
energy coincidence events and low energy triggers/anticoincident events.
Events belonging to the first sample (above 100 TeV) are fully 
reconstructed  from both 
experiments and are analysed in terms of primary composition~\cite{macroeascomp}.
The second group contains events in a limited interval
of primary energy (2 TeV to a few tens of TeV) and
then represents an ideal sample of UG muons to test predictions
of different hadronic interaction models~\cite{macroeasint}. This provides
a unique link between the EAS  and the CR direct measurements.

\section{CONCLUSION}

Large UG detectors showed capability to detect rare signals difficult
to be observed in other experiments. This can be easily recognized
looking at the extraordinary variety of physics topics investigated
by these detectors, including those discussed
in this paper and other fundamental studies covered by other papers
({\it e.g.}, on solar and atmospheric neutrinos, and UHE Neutrino Astrophysics).
The conclusion of this review is that the prospects for very interesting 
developments in the near future are excellent.

Apart from new results from experiments already in operation
(in particular, Super-K will play a leading role among them), we wait  
remarkable progress in two main directions:
\begin{itemize}
\item Among natural sources, solar neutrinos will be captured by a new
generation of detectors, like SNO and BOREXino. These new experiments
are expected to enrich our knowledge about the Sun and neutrinos
originating from it.
\item New `artificial' sources will illuminate UG detectors. At this Workshop
reports have been given about Long Baseline neutrino beams towards
three major UG sites: i)~from KEK PS to Super-K (K2K~\cite{K2K}), 
ii) from Fermilab Main Injector
to Soudan (MINOS~\cite{Barish}) and iii) from CERN SPS to Gran Sasso 
(ICARUS~\cite{icaruslbl}, 
NOE~\cite{Campana}, RICH~\cite{Ypsilantis},...). This is a newly growing field
which is crucial for the understanding of the atmospheric neutrino 
`anomaly'\footnote{
The recent results of Super-K on atmospheric 
neutrinos~\cite{Kearns}, confirming the `anomaly', but at lower $\Delta m^2$
and the new results of the Chooz experiment~\cite{Chooz}, excluding 
large amplitude $\bar{\nu_e} \leftrightarrow \bar{\nu_x}$ oscillations in 
the same $\Delta m^2$ range, render LBL searches considerably more 
troublesome than expected at the time of the Workshop.}.
\end{itemize}

\bigskip

\leftline{\bf Acknowledgments}
\smallskip

I wish to thank G. Battistoni, F. Cei, W. Fulgione, 
M. Goodman, S. Kasahara, E. Kearns, S. Mikheyev, T. Montaruli and  O. Palamara
for helpful discussions and for kindly providing material for my talk as
well as for my paper.

\end{document}